\documentclass[preprint,showpacs,preprintnumbers,amsmath,amssymb,floatfix]{revtex4}
\usepackage{graphicx}
\usepackage{dcolumn}
\usepackage{bm}
\usepackage{longtable}
\begin{document}

\title{ Strain driven magnetism in thin films of $LaCoO_{3}$ } 
\author{Kapil Gupta and Priya Mahadevan}
\affiliation{Department of Materials Science, 
S.N. Bose National Centre for Basic Sciences, JD Block, Sector III, Salt Lake, Kolkata 700098, India }
\date{\today}

\begin{abstract}
The electronic structure of epitaxial films on $LaCoO_{3}$(LCO) has been studied within first principle electronic structure calculations. A spin state transition is found to take place as a function of lattice strain which freezes  in the intermediate spin state for non zero strain. In contrast to earlier speculations this is found to arise from substrate strain alone and angle variations are small. The intermediate spin state also stabilizes ferromagnetism in the ground state. The anomalous temperature dependence of the $X$ ray absorption spectra for films of LCO on LCO and its absence in films of LCO on $(LaAlO_{3})_{0.3}(SrAlTaO_{6})_{0.7}$(LSAT) is also explained.
 
\end{abstract}

\maketitle


3$d$ transition metal oxides exhibit a wide range of electronic and magnetic properties
as a result of coexisting spin, orbital and lattice degrees of freedom, all of which requiring
treatment at the same footing \cite{fujimori}. In recent times there has been a resurgence of interest in these 
materials with the specific aim of controlling their properties with external parameters such as 
strain, electric field etc \cite{strain}. This is with the view of using them as new generation electronic 
components. A material that has been studied in this context is $LaCoO_{3}$(LCO). Bulk LCO \cite{bulk} shows 
interesting temperature dependent properties and the aim has been to examine how one can modify the 
properties externally. Substrate strain \cite{fuchs} has been found to be a useful parameter to control 
the magnetism in the LCO overlayers. More recently LCO films have been grown on a piezoelectric substrate 
\cite{efield}. An electric field has been used to modify the substrate lattice constants, and therefore 
the strain and consequent magnetism in the LCO overlayers. However the exact mechanism by which the 
magnetic state is altered is not very clear.

 Bulk $LaCoO_{3}$ at low temperature is found to exhibit a low spin state with an electronic 
configuration of $t_{2g}^{6}e_{g}^{0}$ on Co \cite{bulk,goodenough}. 
Susceptibility data \cite{susceptibility} show a maximum at 90K 
followed by a Curie Weiss like 
decrease at higher temperatures. This has been interpreted as arising 
from a spin state transition taking place as a function of temperature. 
The nature of the transition, however is still very controversial with 
the debate \cite{korotin,bulk,debate} being whether it is a low spin to high spin transition or 
one to an intermediate spin state. This crossover takes place as a result of a delicate interplay 
between the crystal field splitting and the intratomic exchange interaction.
The temperature dependence is brought about by the 
dependence of the crystal field splitting on the bondlength \cite{bondlength} which in turn 
changes with temperature. This naturally suggests epitaxial strain as an 
alternate handle to tune the spin state transition and therefore change 
the magnetism.

Recently Fuchs {\it et al} \cite{fuchs} have grown thin films of 
LCO on different substrates. Ferromagnetism has been 
found with a $T_{c}$(Curie temperature) of 85K on 
$(LaAlO_{3})_{0.3}(SrAlTaO_{6})_{0.7}$ (LSAT) substrate, 
in addition to a strain dependent $T_{c}$. The origin of the 
ferromagnetism is however not clear, with a significant role played 
by the rotation of the CoO$_{6}$ octahedra being offered as one of 
the reasons \cite{fuchs}. To address this issue we have considered tetragonal unit 
cells of LCO where the inplane and out of plane lattice constants have been kept fixed to the experimental values \cite{fuchs}. We however 
allowed for a rotation of the CoO$_{6}$ octahedra which is a commonly observed lattice distortion in perovskite oxides \cite{mizokawa}and optimized the total energy as a function of the 
angle. The paper by Fuchs {\it et al} \cite{fuchs} speculated that the Co-O-Co angle strongly 
deviated from $180^{0}$ for thin films grown on $LaAlO_{3}$(LAO) and LCO substrate. However, the angles they speculated reached a 
value close to $180^{0}$ in the films grown on LSAT and $SrTiO_{3}$(STO). Thus the strong angle dependence 
of strain drove the spin state transition and hence gave rise to ferromagnetism. Contrary to their 
speculations we find that there is a very slight angle dependence of the strained films. This therefore 
cannot be the reason for the spin state transition observed by us in our calculations. We therefore 
conclude that it is the strain induced pseudo tetragonal structure which is responsible for the spin 
state transition. Further as support to the model proposed by us, we are able to explain the temperature 
dependence of the $X$ ray absorption spectra within our calculations. Experimentally it was found that the 
films of LCO grown on LCO showed a strong temperature dependence of the O K edge as well as the Co L edge $X$ ray 
absorption spectra\cite{pinta}. The L edge $X$ ray absorption spectra of the transition metal atom is 
strongly sensitive to crystal field and spin state effects. Hence the temperature dependence has been 
compared with cluster calculations for CoO$_{6}$ clusters and interpreted as arising from spin 
state transitions. LCO films grown on LSAT substrate however did not show any significant temperature 
dependence. Comparing the total energies obtained by us from calculations for different magnetic states, 
we find that for LCO on LCO the nonmagnetic solution as well as  the other magnetic solutions 
lie very close in energy. The magnetic solutions correspond to an intermediate spin state. Temperature effects we show change the relative concentrations of low spin and intermediate spin states. Hence explaining the temperature dependence of the spectra. However as the substrate strain is varied, the intermediate spin state gets frozen 
in as the ground state, the low spin state lies much higher in energy and hence there is no temperature dependence.

We have performed {\it ab initio} calculations for the electronic structure of thin films of LCO 
using a plane wave pseudopotential implementation of density functional theory as implemented 
in VASP\cite{vasp}. PAW potentials\cite{paw} have been used, in addition to the GGA approximation to 
the exchange part of the functional. A tetragonal unit cell was considered by us where the lattice 
constants were fixed at the experimental values \cite{fuchs}. The effect of the substrate was included by fixing the 
inplane lattice constant to that of the substrate. We also included a GdFeO$_{3}$ type rotation of 
the octahedra which is normally observed in perovskite oxides \cite{mizokawa} and the total energy was minimized 
as a function of the angle. A k point mesh of 4x4x4 was considered for the total energy calculations, 
but increased to 8x8x8 for the density of states calculations using the tetrahedron method. 
Spheres of radii 1.3\AA, 1.2\AA and 1.2\AA were considered on La, Co and O for evaluating 
the magnetic moment and the orbital projected density of states. 

In Fig 1. we have plotted the variation of the magnetic moment on the Co site as a function of the inplane lattice parameter. For LCO films on LCO we show both the nonmagnetic moment as well as the moment for the structure exhibiting the ferromagnetic state. 
The dependence on the inplane latice constant is non monotonic. It should be noted that LCO exhibits a pseudotetragonal unit cell for all values of the substrate lattice constant except for LCO on LCO where it is pseudocubic. Further LCO on LAO represents compressive strain while LCO on LSAT and STO represents tensile strain. Examining the total energies given in Table 1, we find that the ferromagnetic state is the ground state in every case except for LCO on LCO where all magnetic states as well as the nonmagnetic state lie very close to each other in energy.

The Co-O-Co angle variations for the ground state structure as a function of the in and out of plane lattice constant are given in Table 2. Earlier reports suggest that a change in the angle drives the spin state transition. The inplane angle changes are small and can not explain the stabilization of the ferromagnetic state for finite strain. The out-of-plane Co-O-Co angles are found to decrease with the strain in contrast to earlier speculations where angles were expected to approach 180$^{0}$ for films of LCO on LSAT/STO. Hence it is primarily the change in bondlengths as a result of substrate strain which drives the system into the ferromagnetic state.

The question we asked next was what was the spin state stabilized in the ferromagnetic state. To address this we have plotted the Co $d$ projected up and down spin partial density of states for $t_{2g}$ and $e_{g}$ symmetries in Fig 2, for LCO on LSAT. In the tetragonal case there is a further splitting of the $t_{2g}$ orbitals into doubly degenerate $d_{xz}$ and $d_{yz}$ as well as singly degenerate $d_{xy}$ orbital. Similarly there is a splitting of the $e_{g}$ orbitals into singly degenerate $d_{x^{2}-y^{2}}$ and $d_{z^{2}-y^{2}}$ orbitals. However these splittings are small and so we choose to still discuss the electronic structure in terms of the nomenclature valid for the cubic case. From the density of states(Fig 2a) we find that the up spin $t_{2g}$ states are fully occupied while the down spin $t_{2g}$ states are partially occupied. In addition the up spin $e_{g}$ states are partially occupied(Fig 2b). For a high spin configuration we would have the $t_{2g}$ and $e_{g}$ up spin states fully occupied before the $t_{2g}$ down spin states are filled. Hence an intermediate spin state($t_{2g\uparrow}^{3}t_{2g\downarrow}^{2}e_{g\uparrow}^{1}$) is stabilized on Co for LCO on LSAT. This is seen to be the case even for LCO on LCO for the ferromagnetic case (Fig 3a). The partial occupancy of the $t_{2g}$ down spin levels as well as the $e_{g}$ up spin levels favors a ferromagnetic state as the ground state for LCO on LSAT and STO.

Recent $X$ ray absorption experiments\cite{pinta} carried out at the Co $L_{2,3}$ edge for the epitaxial films showed a strong temperature dependence for LCO on LCO. This was absent for LCO films grown on LSAT. They interpreted the results in terms of a spin state transition taking place as a function of temperature for LCO on LCO films. Our total energy calculations for magnetic and nonmagnetic solution indicate that the solutions lie very close in energy for LCO on LCO while the difference is large in all other cases considered by us. The effect of temperature is simulated by us by considering pseudocubic unit cells with a uniform expansion of the unit cell volume. We have computed the total energies for different magnetic configurations and we find that for an expansion of 1$\%$ of the lattice constant, the ferromagnetic state gets stabilized by 60 meV. Thus as a function of temperature the relative weight of intermediate and low spin state in the ground state wavefunction changes giving rise to the temperature dependence. Plotting the Co $d$ projected partial density of states for LCO on LCO considering the nonmagnetic and ferromagnetic solutions we find that the density of states are very different in the two cases(Fig3). The nonmagnetic solution(Fig 3a) has the low spin stabilized while the ferromagnetic solution(Fig 3b) has the intermediate spin stabilized. The partial density of states are consequently different in the two cases. Although multiplets are important in the description of the $X$ ray absorption spectra at the $L_{2,3}$ edge, these results indicate that the final states are different and can therefore explain the temperature dependence seen in experiment.

It is for this reason that we choose to explain the O K edge $X$ ray absorption spectra where initial state core hole effects as
 well as multiplet effects are not important. The experimental spectrum corresponds to transitions from the oxygen 1s level to 
the unoccupied oxygen states with 2p character. Hence the experimental spectrum
may be compared with the broadened O $p$ partial density of states, with the 
broadening account for instrumental resolution among other 
effects. The results of such a comparison 
are shown in Fig. 4 considering the calculated O $p$ density of states for the
ferromagnetic case as well as the nonmagnetic case. There is transfer of 
spectral weight in the low energy region from the nonmagnetic to the ferromagnetic spectral function. A similar transfer of spectral weight is seen in the 
high energy region around 3.0-4.0 eV. This could explain the trend seen 
in experiments for LCO on LCO as a function of temperature.

In the results discussed till now we have analyzed the results from bulk calculations for $LaCoO_{3}$ where the effect of the substrate is taken in defining the inplane lattice constants. The electronic and magnetic structurre could strongly deviate at the interface as well as at the surface. In order to analyze this we have considered films of $LaCoO_{3}$ grown on STO substrate consisting of 16 layers of LCO grown in a symmetric slab arrangement on STO. At the surface as well as at the interface we find a significant reduction in the moment from bulk-like values. However no significant moment or occupancy of the Ti layers is found.
 
We have carried out {\it ab initio} electronic structure calculations for epitaxial films of $LaCoO_{3}$ grown on various substrates. We find that the intermediate spin is frozen in for the cases in which a pseudo tetragonal structure is stabilized and the films are subject to compressive/tensile strain. The stabilization of the intermediate spin state also makes the ferromagnetic state to have lowest energy. LCO on LCO is found to have a pseudocubic structure. Total energy calculations reveal that the nonmagnetic and magnetic solutions lie close in energy for the ground state lattice constant with the energy difference changing for a uniformly expanded case, thus explaining the temperature dependence observed in $X$ ray absorption spectra.

We thank the Department of Science and Technology, Government of India for financial support. We also thank Ashis Kumar Nandy for careful reading of the manuscript.

\renewcommand

\begin{figure}
\includegraphics[width=5.5in,angle=270]{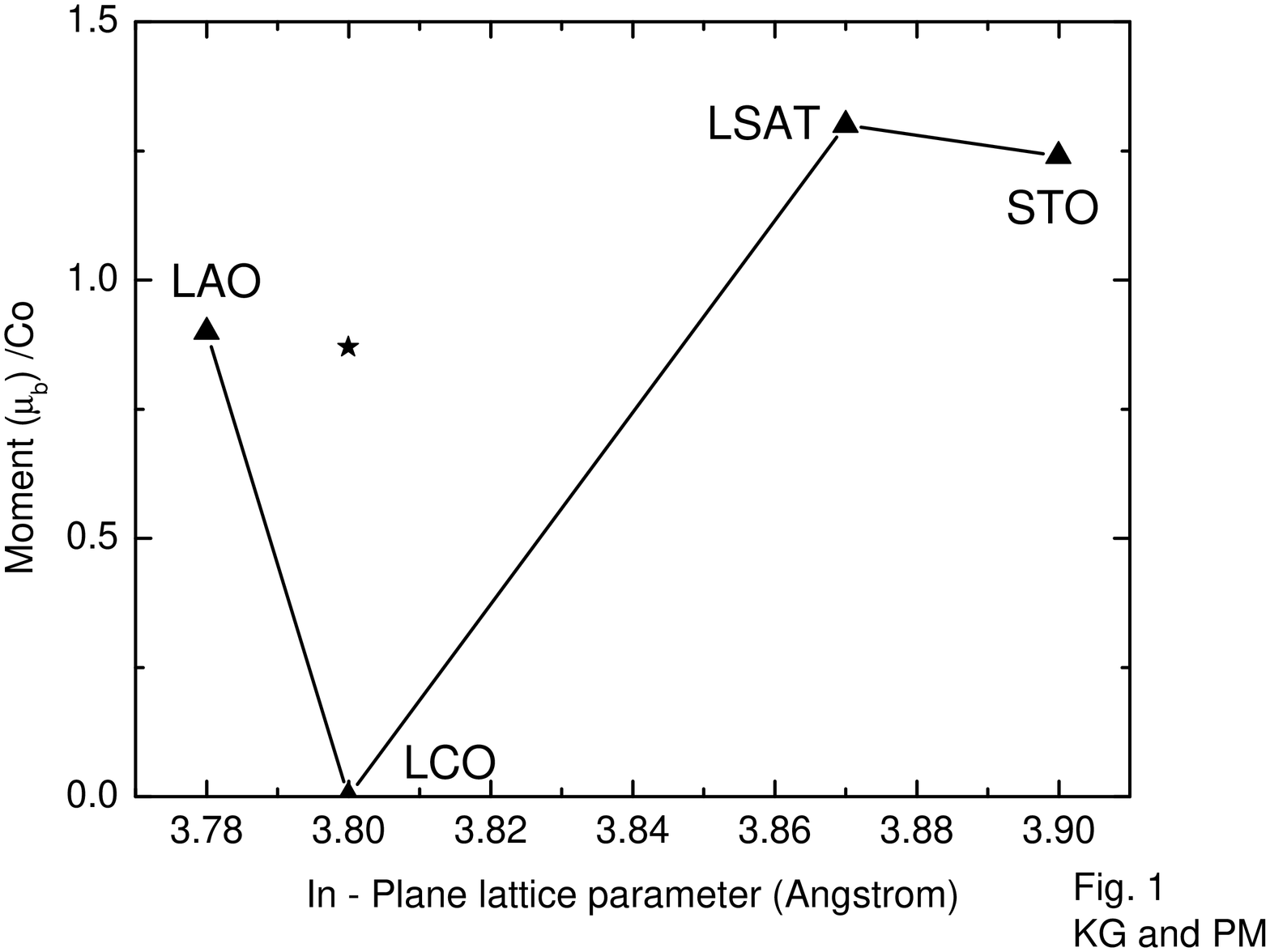}
\caption{ The variation of the magnetic moment(triangles) on the Co site as a function of the inplane lattice parameter for the ground state. The nature of the substrate has also been indicated. For LCO on LCO we also provide the moment of the ferromagnetic state(star) which lies close in energy to the nonmagnetic states.}

\end{figure}

\begin{figure}
\includegraphics[width=5.5in,angle=270]{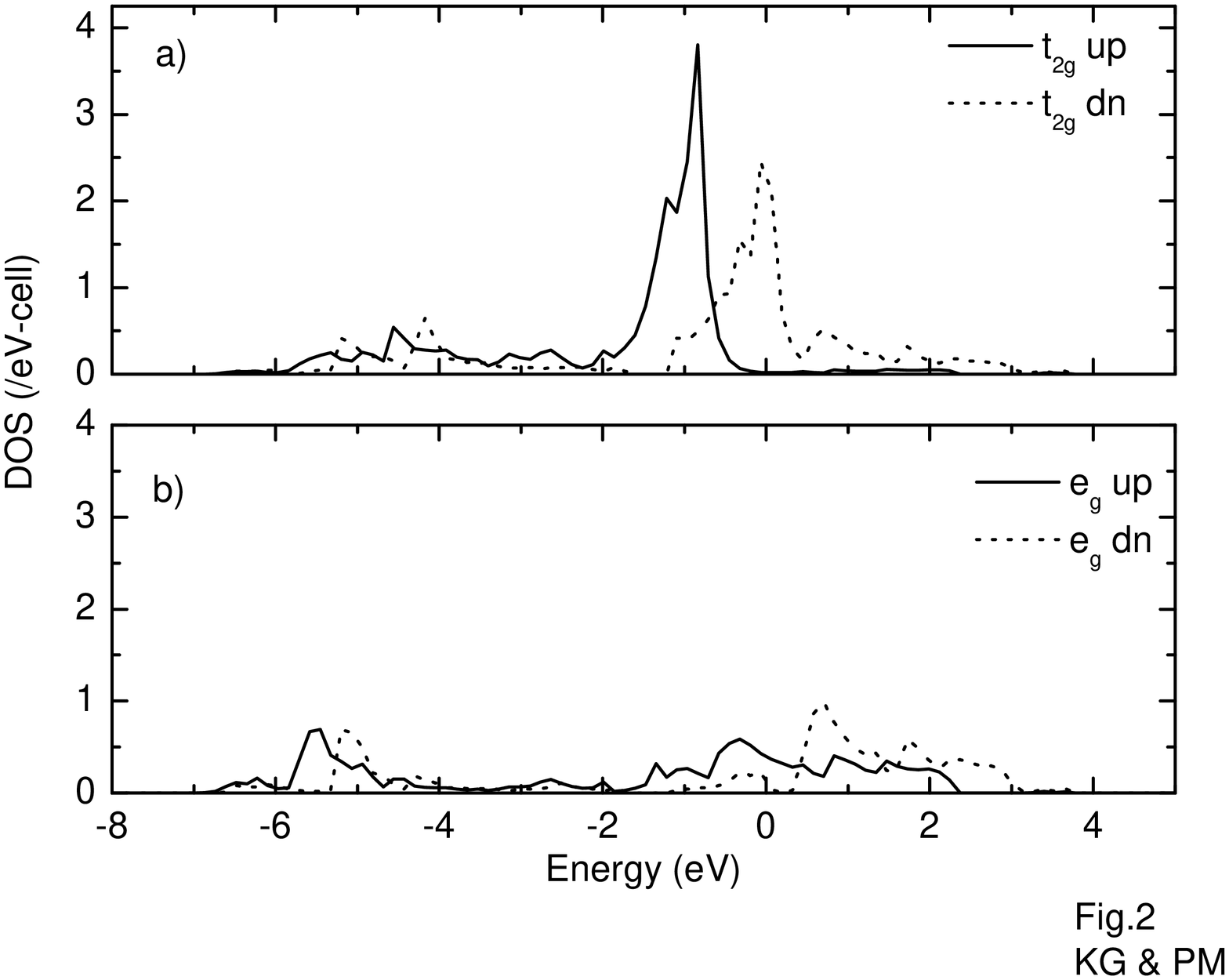}
\caption{Co $d$ projected up and down spin partial density of states for $t_{2g}$ and $e_{g}$ symmetries for LCO on LSAT }
\end{figure}

\begin{figure}

\includegraphics[width=5.5in,angle=270]{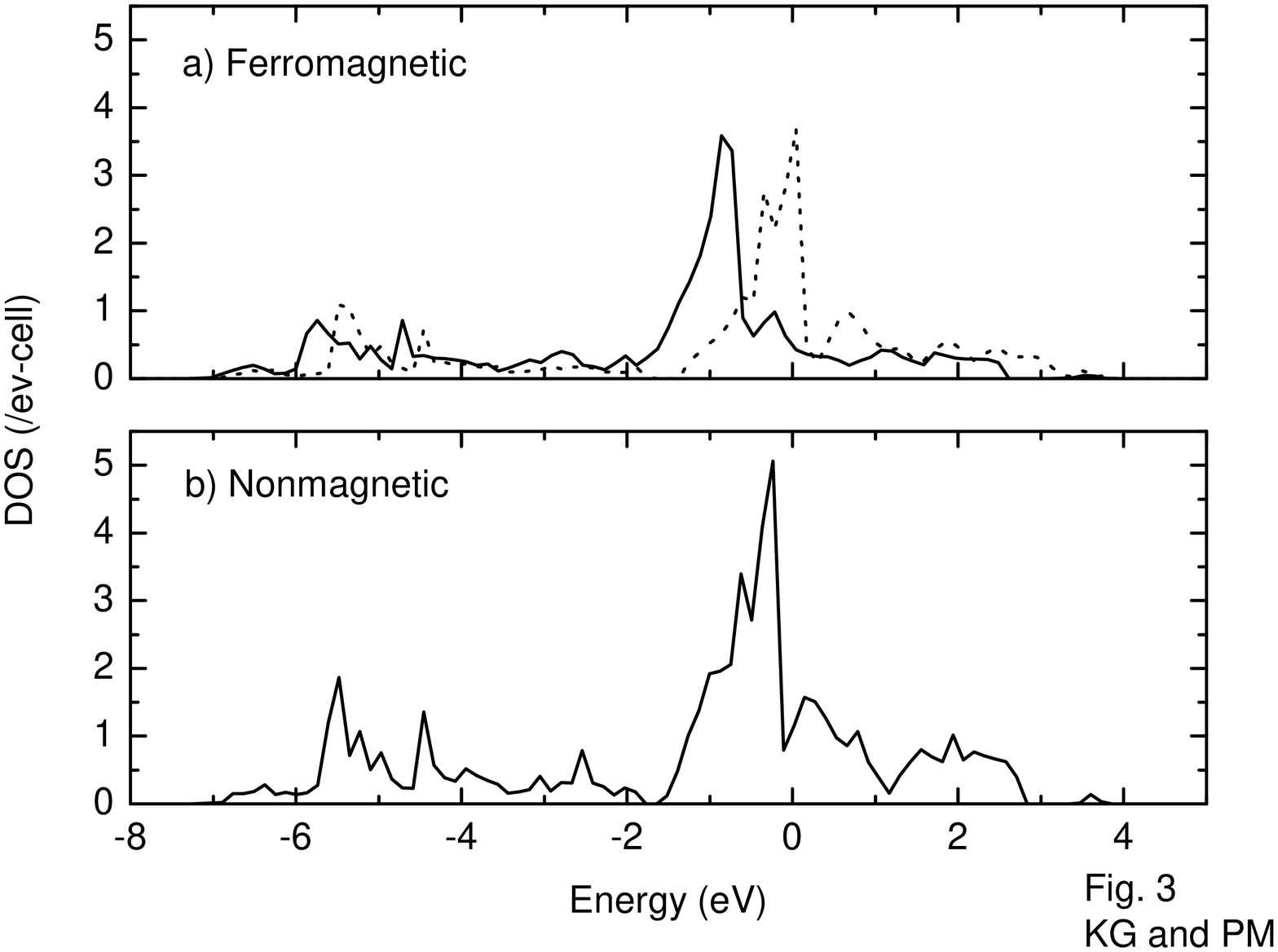}
\caption{Co $d$ projected partial density of states for (a)ferromagnetic case and (b)nonmagnetic case for films of LCO grown on LCO}

\end{figure}

\begin{figure}

\includegraphics[width=5.5in,angle=270]{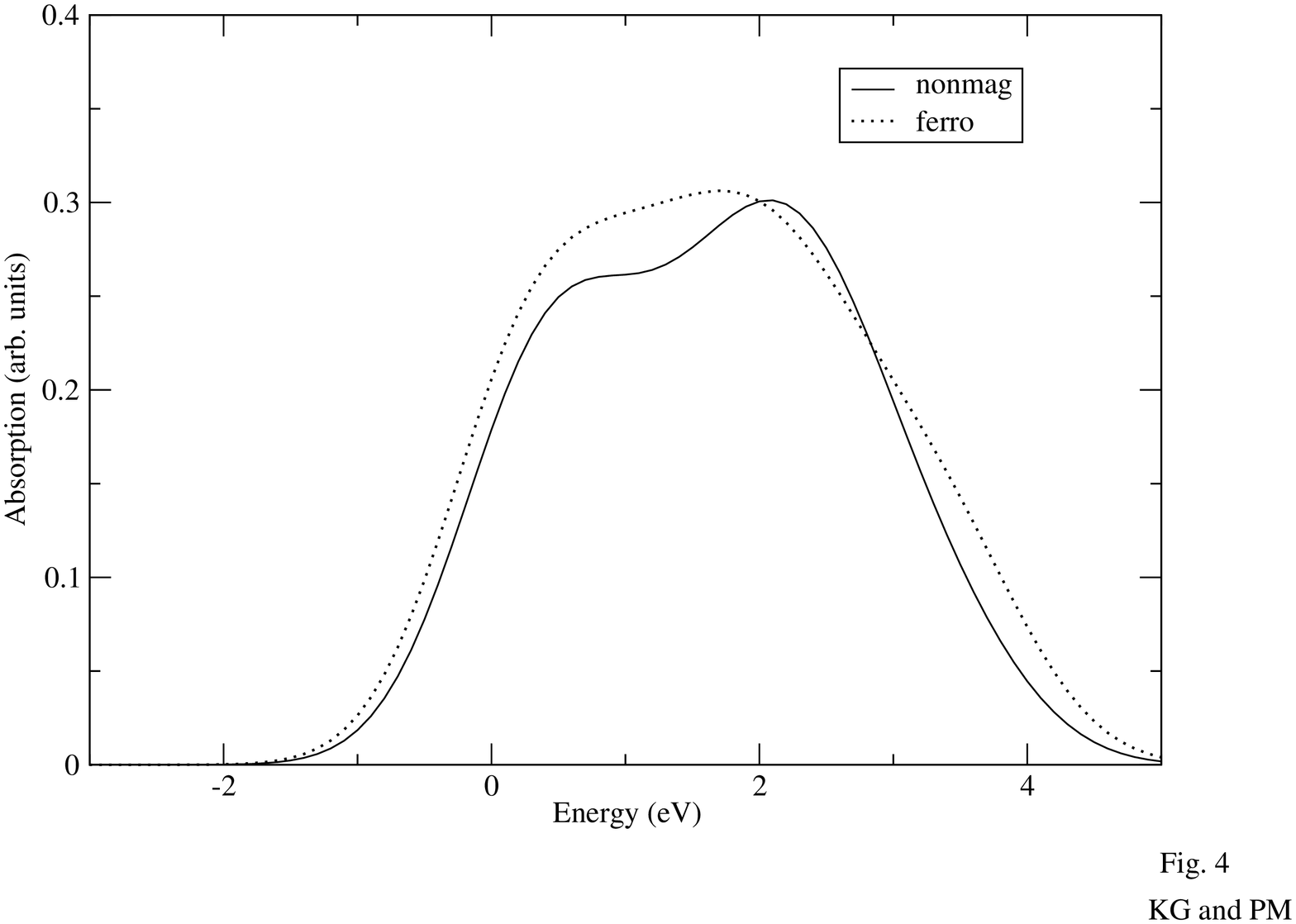}
\caption{Calculated O K edge xray absorption spectra for nonmagnetic (solid line)as well
as ferromagnetic cases (dashed line).}
\end{figure}

\newpage
\begin{table}
\caption
{Energies in eV for 4 formula units of LCO grown on different substrates for different magnetic configurations.}
\begin{tabular}{|c|c|c|c|c|c|} 
\cline{2-5}
\hline 
\hline
& \multicolumn{1}{c} {LCO/LCO} \vline & \multicolumn{1}{c} {LCO/LAO} \vline & \multicolumn{1}{c} {LCO/LSAT} \vline & \multicolumn{1}{c} {LCO/STO} \vline\\
\hline
Non-magnetic & -152.761 & -152.742 & -152.875 & -152.793 \\

A-type   & -152.746 & -152.719 & -152.827 & -152.845 \\

C-type   & -152.761 & -152.736 & -152.862 & -152.793 \\

Ferro-magnetic & -152.742 & -152.781 & -152.953 & -152.939 \\

\hline\hline
\end{tabular}
\end{table}

\begin{table}
\caption
{Co-O-Co angle in x- y- and z- directions for LCO grown on different substrates. In- and out- of plane
lattice constants are also given.}
\begin{tabular}{|c|c|c|c|c|c|}

\hline \hline
 & \multicolumn{2}{c} {Lattice Constant}  \vline &  \multicolumn{3}{c} {Co-O-Co Angle} \vline  \\
\cline{2-6}

Substrate &  Inplane & Out of plane &  x direction & y direction & z direction   \\
\hline
LAO & 3.78 & 3.87 & 163.3 & 163.3 & 160.9 \\
LCO & 3.8 & 3.8 & 162.2 & 162.2 & 158.3 \\
LSAT & 3.87 & 3.8 & 163.2 & 163.2 & 157.6 \\
STO & 3.9 & 3.79 & 163.1 & 163.1 & 156.7 \\
\hline\hline
\end{tabular}
\end{table}


\newpage

\begin{references}

\bibitem{fujimori}
Masatoshi Imada, Atsushi Fujimori, and Yoshinori Tokura, Rev. Mod. Phys. {\bf 70}, 1039 (1998).
\bibitem{strain}
M. Dawber, K. M. Rabe, and J. F. Scott, Rev. Mod. Phys. {\bf 77}, 1083 (2005); 
V. Laukhin, V. Skumryev, X. Martí, D. Hrabovsky, F. Sánchez, M. V. García-Cuenca, C. Ferrater, M. Varela, U. L$\ddot{u}$ders, J. F. Bobo and J. Fontcuberta, Phys. Rev. Lett. {\bf 97}, 227201 (2006);
C.Thiele, K. D$\ddot{p}$rr, O. Bilani,1 J. R$\ddot{o}$del and L. Schultz, Phys. Rev. B {\bf 75}, 054408 (2007);
A.D. Rata, A. Herklotz, K. Nenkov, L. Schultz and K. D$\ddot{o}$rr, Phys. Rev. Lett. {\bf 100}, 076401 (2008).
\bibitem{bulk} 
S.R. Barman and D.D. Sarma, Phys. Rev. B {\bf 49}, 13979 (1994);
M. Abbate, R. Potze, G.A. Sawatzky and A. Fujimori, Phys. Rev. B {\bf 49}, 7210 (1994)
T. Saitoh, T. Mizokawa, A. Fujimori, M. Abbate, Y. Takeda, and M. Takano, Phys. Rev. B {\bf 55}, 4257 (1997). 
\bibitem{fuchs}
D. Fuchs,  C. Pinta, T. Schwarz, P. Schweiss, P. Nagel, S. Schuppler, R. Schneider, M. Merz, G. Roth, and H. v. L$\ddot{o}$hneysen, Phys. Rev. B {\bf 75}, 144402 (2007);
D. Fuchs, E. Arac, C. Pinta, S. Schuppler, R. Schneider, and H. v. L$\ddot{o}$hneysen, Phys. Rev. {\bf 77}, 014434 (2008).
\bibitem{efield}
A. Herklotz, A. D. Rata, L. Schultz and K. D$\ddot{o}$rr, arXiv: 0805.0991v1 
[cond-mat.str-el].
\bibitem{goodenough}
J. Q. Yan, J. S. Zhou and J. B. Goodenough, Phys. Rev. 
B {\bf 69}, 134409 (2004).
\bibitem{susceptibility}
C. Zobel, M. Kriener, D. Bruns, J. Baier, M. Gr$\ddot{u}$ninger, T. Lorenz, P. Reutler and A. Revcolevschi, Phys. Rev. B {\bf 66}, 020402(R), (2002).
\bibitem{debate}
M.W. Haverkort, Z. Hu, J. C. Cezar, T. Burnus, H. Hartmann, M. Reuther, C. Zobel, T. Lorenz, A. Tanaka, N. B. Brookes, H. H. Hsieh, H. J. Lin, C. T. Chen, and L. H. Tjeng, Phys. Rev. Lett. {\bf 97}, 176405 (2006);
A. Podlesnyak, S. Streule, J. Mesot, M. Medarde, E. Pomjakushina, K. Conder, A. Tanaka, M. W. Haverkort, and D. I. Khomskii, Phys. Rev. Lett. {\bf 97}, 247208 (2006);
R.F. Klie,  J. C. Zheng, Y. Zhu, M. Varela, J. Wu and C. Leighton, Phys. Rev. Lett. {\bf 99}, 047203 (2007).
\bibitem{korotin}
M. A. Korotin, S. Yu. Ezhov, I. V. Solovyev, V. I. Anisimov, D. I. Khomskii and G. A. Sawatzky, Phys. Rev. B {\bf 54}, 5309(1996).
\bibitem{mizokawa}
T. Mizokawa and A. Fujimori, Phys. Rev. B {\bf 54}, 5368(1996).
\bibitem{bondlength}
D.M. Sherman, in Advances in Physical Geochemistry, edited by S.K. Saxena
(Springer Verlag, Berlin 1988).
\bibitem{pinta}
C. Pinta, D. Fuchs, M. Merz, M. Wissinger, E. Arac, A. Samartsev, P. Nagel, S. Schuppler, arXiv:0807.3424v1 [cond-mat.str-el].
\bibitem{vasp}
G. Kresse, and J. Furthm$\ddot{u}$ller, Phys. Rev. B. {\bf 54}, 11169 (1996); G.~Kresse, and J.~Furthm$\ddot{u}$ller, Comput. Mat. Sci. {\bf 6}, 15 (1996).
\bibitem{paw}
G. Kresse, and J. Joubert,Phys. Rev. B {\bf 59}, 1758 (1999).


\end{references}
\end{document}